\documentstyle[12pt]{article}
\begin{document}
\begin{center}
{\large\bf
Extended non-chiral quark models confronting QCD}\footnote{Talk given at the 
International Workshop on Hadron Physics, 
Effective Theories of Low Energy QCD, Coimbra, Portugal, September 1999.}\\

\vspace{1cm}

{\bf
A.A.Andrianov}$^{*,\dagger}$\footnote{Supported by Grant
GRACENAS 6-19-97 and by Generalitat de Catalunya,
Grant PIV 1999.} and 
{\bf V.A.Andrianov}$^{\dagger}$\footnote{Supported by Grant
RFBR 98-02-18137,  Travel Grant of Russian Academy of Science
and by Funds of the Workshop Hadron 99, Coimbra.}\\
\vspace{0.5cm}

$^*$Departament d'ECM, Universitat de Barcelona\\
 08028 Barcelona, Spain\\
$^{\dagger}$Department of Theoretical Physics,
St.-Petersburg State University,\\
198904 St.-Petersburg, Russia
\end{center}

\vspace{1cm}

\begin{abstract}
We discuss the   low  energy effective   action  of QCD in the   quark
sector. When it is built at  the CSB (chiral   symmetry breaking) scale  by
means of perturbation  theory it has the  structure
of a generalized Nambu-Jona-Lasinio (NJL) model  with CSB due to
attractive  forces in the scalar channel.
We  show   that   if the  lowest  scalar    meson state  is
sufficiently  lighter than the heavy pseudoscalar $\pi'$
then   QCD favors a low-energy   effective  theory in which 
 higher
dimensional  operators  (of   the Nambu-Jona-Lasinio  type)  
are dominated and relatively  strong.  A light scalar quarkonium ($m_\sigma
= 500 \div 600$ MeV) would provide an 
evidence in favor to this NJL mechanism.

Thus the non-chiral Quasilocal Quark Models (QQM)
in the dynamical 
symmetry-breaking regime are considered as approximants for low-energy action
of QCD. 
In the mean-field (large-$N_c$) approach 
the equation on critical coupling surface is
derived. The mass spectrum of
scalar and pseudoscalar excited states is calculated in leading-log 
approach which is compatible with the truncation of the QCD effective
action with few higher-dimensional operators. 
The matching to QCD based on the Chiral Symmetry
Restoration sum rules is performed and it helps to select out the
relevant
pattern of CSB as well as to enhance considerably the predictability of 
this approach.
\end{abstract}
\vfill
\noindent
hep-ph/9911383
\newpage

\section*{INTRODUCTION: DEFINITION OF QQM}
The low energy effective
action of QCD in the quark sector
has a qualitatively different structure
depending on whether it is built
at the CSB scale by means of perturbation theory or below the CSB scale
when the major chiral symmetry breaking effect - the formation of
light pseudoscalar mesons - is implemented manifestly.
In the first case the models \cite{andr1993} extend the Nambu-Jona-Lasinio
one \cite{nambu1961} (see the reviews
\cite{volk1984,meissn1988,voglweise1991,klevan1992,andr1992,bijnbruraf,hatsud1994,ebereinvol1994,bijnen1996})
  with chiral symmetry broken due to strong 
attractive 4-fermion
forces in the color-singlet scalar channel.
In the second case the
resulting model\cite{andresp1998} is a generalization  
of the  chiral quark model
with a built-in constituent quark mass
and the  non-linear realization of chiral symmetry.
 This type of
QCD effective action is not discussed in our talk although it  may be
more relevant \cite{andresp1998,andresp1999,andresptar1999} 
if the quarkonium scalar meson is sufficiently heavy
\cite{oset1998,schechter1999}.

The quasilocal approach of \cite{andr1993} (see also 
\cite{andr1995,pallan1995,andryudi1996})
represents a systematic extension of the NJL model towards the complete
effective action of QCD where many-fermion
vertices with
derivatives are incorporated with the manifest chiral symmetry of
interaction,  motivated by the soft momentum
expansion of the perturbative QCD effective action.
For sufficiently strong couplings,
the new operators promote the formation of additional
scalar
and pseudoscalar states. These models allow an extension of the
linear $\sigma$ model provided by the NJL model,  with the pion being
a broken symmetry partner of the
lightest scalar meson just as before, and with excited pions and scalar
particles coming in pairs. In particular, when only scalar and pseudoscalar
color-singlet channels are examined and dynamical quark masses are supposed to
be sufficiently smaller than the CSB cutoff one may derive 
the minimal two-channel
lagrangian of the QQM in the separable form \cite{andr1993,andryudi1996}:
\begin{eqnarray}
{\cal L}^{QQM}&=&\bar q i /\!\!\!\partial q + {\cal L}^{I};\nonumber\\
{\cal L}^{I}&=&
\frac{1}{4 N_f N_{c}\Lambda^{2}}\sum_{k,l=1}^{2}a_{kl}
\left[\bar q f_k(\hat s)q \,\bar q f_l(\hat s) q
- \bar q f_k(\hat s) \tau^a \gamma_5 q \, 
\bar q f_l(\hat s)\tau^a \gamma_5 q\right], \label{lagQQM}
\end{eqnarray}
where  $/\!\!\!\partial \equiv \gamma^\mu \partial_\mu$, $a_{kl}$ represents a
symmetric matrix of real coupling constants and polynomial formfactors
are chosen as follows:
\begin{equation}
f_{1}(\hat s) =2-3 \hat s \,; \qquad
f_{2}(\hat s) = -\sqrt{3} \hat s\,; \qquad \hat s \equiv
-\frac{\partial^{2}}{\Lambda^{2}}\,.
\end{equation}
As this model interpolates the low-energy QCD action it is supplied with
the cutoff $\Lambda \sim 1$ GeV which bounds virtual quark momenta in
quark loops. We restrict ourselves with consideration of two-flavor case,
thus $\tau_a$ denote Pauli matrices.

A somewhat different, nonlocal approach to describe excited meson states was
developed in \cite{volkweiss1997,volkebert1998}

\section*{EFFECTIVE POTENTIAL AND MESON SPECTRUM}
For  strong four-fermion coupling constants $a_{kl} \sim 8\pi^2 \delta_{kl}$
the lagrangian
(\ref{lagQQM}) reveals the phenomenon of dynamical chiral symmetry 
breaking. This phenomenon can be described
with the help of the effective potential for the
attractive scalar channel where scalar mesons arise as composite
states. Indeed its
non-trivial minimum 
gives rise to a dynamical quark mass and the perturbative fluctuations
around this minimum characterize the mass spectrum of meson states.
To derive the required effective potential one should bosonize the
quark action, i.e. incorporate auxiliary bosonic variables:
$\sigma_k \sim i \bar q f_k(\hat s)q ,\quad 
\pi_k^a \sim \bar q f_k(\hat s) \tau^a \gamma_5 q\quad $ and
integrate out fermionic degrees of freedom.
At the first step we introduce the bosonic variables in two channels:
\begin{equation}
{\cal L}_{I}= \sum_{k=1}^{2} i \bar q\left(\sigma_k + 
i \gamma_5\pi_k^a\tau^a\right)  f_k(\hat s) q
+ N_f N_{c}\Lambda^{2}\sum_{k,l=1}^{2}
\left(\sigma_k a_{kl}^{-1}\sigma_l + \pi_k^a a_{kl}^{-1}\pi_l^a\right).
\label{bosonization}
\end{equation}
Let us parametrize the matrix of coupling constants in a close vicinity
of tricritical point: 
\begin{equation}
 8\pi^2a_{kl}^{-1} = \delta_{kl} - \frac{\Delta_{kl}}{\Lambda^2},\qquad
|\Delta_{kl}| \ll \Lambda^2.
\label{couplQQM}
\end{equation}
The last inequality turns out to be equivalent to require
the dynamical mass to be essentially less than the cutoff.

After integrating out the quark fields one comes to the
bosonic effective action ${\cal W}(\sigma_k, \pi_k^a)$ 
and therefrom,  for constant meson variables,
to the effective potential:
\begin{eqnarray}
 V_{eff}&=&
\frac{N_{c}N_f}{8\pi^2}\Biggl(
-\sum_{k,l=1}^{2}\sigma_{k}\sigma_{l}\Delta_{kl} - 
 (\pi_2^a)^2 \Delta_{22} + 
8(\sigma_{1})^{4}\left(
\ln\frac{\Lambda^{2}}{4(\sigma_{1})^2}+\frac12\right)
\nonumber\\
&&
-\frac{159}{8}(\sigma_{1})^{4}-\frac{5\sqrt{3}}{2}
\sigma^{3}_{1}\sigma_{2}
+\frac{9}{4}\sigma^{2}_{1}\sigma^{2}_{2}
+\frac{\sqrt{3}}{2}\sigma_{1}\sigma^{3}_{2}
+\frac{9}{8}(\sigma_{2})^{4}
\nonumber\\
&& +\left(\frac{3}{4}\sigma^{2}_{1}
+\frac{\sqrt{3}}{2}\sigma_{1}\sigma_{2}
+ \frac{9}{4}\sigma^{2}_{2}
\right)(\pi_2^a)^2 + \frac{9}{8}(\pi_2^a)^4
\Biggr)+O\left(\frac{\ln\Lambda}{\Lambda^{2}}\right),
\label{effpotQQM}
\end{eqnarray}
for the fixed direction of chiral symmetry breaking $\pi_1^a = 0$.

The QCD inspired action should not, of course, induce the isospin symmetry
breaking and therefore a non trivial solution for v.e.v is expected to be 
in the scalar channel, $< \pi_2^a > = 0$. It implies the following inequality
to hold:
$$
\frac{3}{4}\sigma^{2}_{1}
+\frac{\sqrt{3}}{2}\sigma_{1}\sigma_{2}
+ \frac{9}{4}\sigma^{2}_{2} > \Delta_{22}.$$

The conditions on extremum of the effective potential (\ref{effpotQQM}),
the mass-gap equations,
\begin{eqnarray}  
\Delta_{11} \sigma_1 +  \Delta_{12} \sigma_2 &=& 16 \sigma_1^{3}
\ln\frac{\Lambda^{2}}{4 \sigma_1^{2}}
- \frac{159}{4} \sigma_1^{3}-\frac{15\sqrt{3}}{4}
\sigma^{2}_{1}\sigma_{2}
+\frac{9}{4}\sigma_{1}\sigma^{2}_{2}
+\frac{\sqrt{3}}{4}\sigma^{3}_{2}\nonumber\\
\Delta_{12} \sigma_1  + \Delta_{22} \sigma_2 &=& 
-\frac{5\sqrt{3}}{4}
\sigma^{3}_{1}
+\frac{9}{4}\sigma^{2}_{1}\sigma_2
+\frac{3\sqrt{3}}{2}\sigma_{1}\sigma^{2}_{2}
+\frac{9}{4}(\sigma_{2})^{3},
\label{massgapQQM}
\end{eqnarray}
 allow to find  the relations between the components of 
dynamical mass function and (reduced) coupling constants $\Delta_{kl}$.
In practice, one uses the v.e.v.'s of scalar fields as input parameters,
in particular, $2\sigma_{1} = m_{dyn} = 200 \div 300$ MeV, and
determines the required $\Delta_{kl}$.

The second variation of effective action in the vicinity of above v.e.v.,
\begin{eqnarray}
\frac{\delta^2{\cal W}}{\delta\sigma_k(p)\delta\sigma_l(p') } &=&
 \frac{N_{c} N_f}{8\pi^2}(A_{kl}^\sigma p^2  + B_{kl}^\sigma) 
\delta^{(4)}(p + p');\nonumber\\
\frac{\delta^2{\cal W}}{\delta\pi_k(p)\delta\pi_l(p') } &=&
 \frac{N_{c} N_f}{8\pi^2} (A_{kl}^\pi p^2  + B_{kl}^\pi) \delta^{(4)}(p + p'),
\label{secondvar}
\end{eqnarray} 
brings both the kinetic terms $\sim A_{kl}^{\sigma,\pi}$ and the mass 
matrix $B_{kl}^{\sigma,\pi}$
which represents the second derivative of the effective potential 
(\ref{effpotQQM}) (see their general 
structure in \cite{andrroden}). 

The kinetic matrices
$A^{\sigma,\pi}_{k,l}$ take the form:
\begin{equation}
 A_{kl}^{\sigma}
 \simeq A_{kl}^{\pi} \simeq \left(\begin{array}{cc}
\left(4\ln\frac{\Lambda^2}{4\sigma^2_1} - \frac{23}{2}\right) &\quad - 
\frac{\sqrt{3}}{2}\\
- \frac{\sqrt{3}}{2} & \quad \frac32
\end{array}\right) .\label{matrixa}
\end{equation}

Let us now display the matrix of second variations $B^{\sigma,\pi}_{k,l}$:
\begin{eqnarray}
B_{11}^{\sigma}&=&
        - 2\Delta_{11}  + 96\sigma_{1}^{2}
        \ln\left({{\Lambda^{2}}\over{4 \sigma_{1}^{2}}}\right)
        -\frac{605}{2}\sigma_{1}^{2}- 15\sqrt{3}\sigma_{1}\sigma_{2}
        + \frac{9}{2} \sigma_{2}^{2}\, ,\nonumber\\   
B_{12}^{\sigma}&=& -2\Delta_{12} -\frac{15\sqrt{3}}{2}\sigma_{1}^{2} +
        9 \sigma_{1}\sigma_{2}+
        \frac{3\sqrt{3}}{2}\sigma_{2}^{2} \, ,\nonumber\\
B_{22}^{\sigma}&=& -2\Delta_{22} +
        \frac92 \sigma_{1}^{2}+ 3\sqrt{3}\sigma_{1}\sigma_{2} +
\frac{27}{2} \sigma_{2}^{2}\, ,\nonumber\\
B_{11}^{\pi}&=&        - 2\Delta_{11}  + 32\sigma_{1}^{2}
        \ln\left({{\Lambda^{2}}\over{4\sigma_{1}^{2}}}\right)
        -\frac{159}{2}\sigma_{1}^{2} - 5\sqrt{3}\sigma_{1}\sigma_{2}
        + \frac{3}{2} \sigma_{2}^{2}\, ,\nonumber\\   
B_{12}^{\pi}&=&-2\Delta_{12}
        - \frac{5\sqrt{3}}{2}\sigma_{1}^{2}+ 3\sigma_{1}\sigma_{2} +
\frac{\sqrt{3}}{2}\sigma_{2}^{2}\, , \nonumber\\
B_{22}^{\pi} &=&-2\Delta_{22} +
        \frac32 \sigma_{1}^{2}+ \sqrt{3}\sigma_{1}\sigma_{2} +
\frac{9}{2} \sigma_{2}^{2} \,. \label{matrixb}
\end{eqnarray}

Their diagonalization 
allows to find the masses of meson states. 
One recovers two states in the scalar
channel and two triplet states in the pseudoscalar one. The lightest
multiplet consists of the massless pion, $m_\pi = 0$ and the NJL
scalar meson, $m_\sigma = 4 \sigma_{1} = 2 m_{dyn}$ in the leading-log
approach.
The masses of heavier mesons depend essentially on the 
pattern of CSB in the vicinity of tricritical point. But  in the
leading-log approach they are approximately equal,
$m_{\sigma'}^2 \sim m_{\pi'}^2 \sim 
-\frac43 \Delta_{22} = {\cal O}(\log\Lambda)$.
The last estimation follows from the mass-gap eqs.(\ref{massgapQQM}). 
A more precise relation takes places:
\begin{eqnarray}
m_{\pi'}^2  &\simeq& -\frac43 \Delta_{22} +  \sigma^{2}_{1}
+\frac{2 \sqrt{3}}{3}\sigma_{1}\sigma_{2}
+ 3 \sigma^{2}_{2};
\nonumber\\
m_{\sigma'}^2 -  m_{\pi'}^2 &\simeq& 2 \sigma^{2}_{1}
+\frac{4 \sqrt{3}}{3}\sigma_{1}\sigma_{2}
+ 6 \sigma^{2}_{2}  > 0.
\label{masrelQQM}
\end{eqnarray}
\section*{CSR RULES}
Let us employ the constraints based
on Chiral Symmetry Restoration (CSR) in QCD at high energies.
We consider two-point correlators of color-singlet 
quark currents in Euclidean space-time,
\begin{equation}
\Pi_C (p^2) = \int d^4x \,\exp(ipx)\
\langle T\left(\bar q\Gamma q (x) \,\, \bar q \Gamma q
(0)\right)\rangle,
\end{equation}
restricting ourselves in this talk with
\begin{equation}
 C \equiv S, P;\qquad \Gamma = i,\, \gamma_5 \tau^a .
\end{equation}
In the chiral limit the scalar correlator $\Pi_S$ and the pseudoscalar one 
 $\Pi_P^{aa}$ approach to each other rapidly as their O.P.E.'s
 \cite{shifm1979,reinrubin1985}
 coincide at all orders
in perturbation theory and, as well, in the non-perturbative,
purely gluonic part \cite{andresp1998,andr1996},
\begin{equation}
\left(\Pi_P^{aa}(p^2)- \Pi_S(p^2)\right)_{p^2 \rightarrow \infty} \equiv
\frac{\Delta_{SP}}{p^4}  + 
{\cal O} \left(\frac{1}{p^6}\right),\qquad\Delta_{SP} \simeq  24 \pi\alpha_s 
< \bar q q >^2,   \label{ChirSymRes}
\end{equation}
where $< \bar q q >$ stands for the quark condensate and 
the vacuum dominance hypothesis \cite{shifm1979} is exploited for the estimation
of four-quark condensates as we follow the
large-$N_c$ limit. Meantime, in the latter limit 
the correlators are well saturated by
narrow resonances,
\begin{equation}
\Pi_P^{aa}(p^2)- \Pi_S(p^2) =
\sum_n \,\left[\frac{Z^P_n}{p^2 + m^2_{P,n}} \, -\,
\frac{Z^S_n}{p^2 + m^2_{S,n}}\right].
\end{equation}
As the difference decreases rapidly, one can
assume that the lower lying resonances will dominate in the above sum.
 
The outcoming CSR rules in the two-channel model (\ref{lagQQM})  read:
\begin{equation}
Z_{\sigma} + Z_{\sigma'} = Z_{\pi} + Z_{\pi'};\qquad
Z_{\sigma} m^2_{\sigma} + Z_{\sigma'} m^2_{\sigma'}
=  Z_{\pi'} m^2_{\pi'} + \Delta_{SP} . \label{constraints}
\end{equation}
The first relation can be fulfilled in the (one-channel) NJL model
which corresponds to the one-resonance ansatz, $Z_{\sigma',\pi'} = 0$,
whereas the last one can be valid only in a two-resonance model,
for the  $\Delta _{SP}$ defined in (\ref{ChirSymRes}) 
(see \cite{andresp1999}).

\section*{CONSTRAINTS ON MESON PARAMETERS}
The relevant correlators and the values of residues $Z_i$ can be found
by variation of the external sources $S_{k}, P^a_{k}$ which couple to the
scalar and pseudoscalar quark densities. 
The structure of the corresponding operators
in the quark lagrangian is completely analogous to the Yukawa vertex in
(\ref{bosonization}). Then the effect of external
sources can be separated by shifting the scalar fields in the quark vertex of
(\ref{bosonization}) and further on by integrating out the quark fields.
As a result the effective action for generating of two-point correlators is 
parametrized in terms of the second variation matrix (\ref{secondvar}):
\begin{eqnarray}
{\cal W}^{(2)} & \simeq& \frac{N_c N_f \Lambda^2}{8\pi^2} 
\sum_{k,l=1}^{2}\left(
S_k \Gamma^{(\sigma)}_{kl} S_l +  P_k^a \Gamma^{(\pi)}_{kl}P_l^a\right),
\nonumber\\
 \Gamma^{(\sigma)}_{kl}& =&\delta_{kl} - 2\Lambda^2 \left(A^{\sigma} p^2 +
 B^{\sigma}\right)^{-1}_{kl}\,;\qquad
\Gamma^{(\pi)}_{kl} =\delta_{kl}  - 2\Lambda^2 \left(A^{\pi} p^2 +
 B^{\pi}\right)^{-1}_{kl}\,. \label{correlator}
\end{eqnarray}
In particular, the strictly local quark densities can be presented as a 
superposition of
two currents:
\begin{equation}
\bar q \Gamma q = \frac12 \left(\bar q f_1\Gamma q 
- \sqrt{3} \bar q f_2 \Gamma q\right);\quad \Gamma \equiv i, \gamma_5 \tau^a.
\label{localden}
\end{equation} 
Respectively, their two-point correlator in the scalar channel,
$\Pi_S(p^2)$ reads,
\begin{eqnarray}
\Pi_\sigma (p^2) &=& - \frac{N_c\Lambda^2}{2\pi^2}
+ \frac{Z_{\sigma}}{p^2 +m^2_{\sigma}} + \frac{Z_{\sigma'}}{p^2 +
m^2_{\sigma'}}\, ;\nonumber\\ 
Z_{\sigma} &\simeq& - \frac{N_c\Lambda^4}{12\pi^2 m^2_{\sigma'}
\ln\frac{\Lambda^2}{4\sigma^2_1}} 
\left[ - 48 \sigma^2_1 \ln\frac{\Lambda^2}{4\sigma^2_1} 
 +  3\Delta_{11} +2\sqrt{3} \Delta_{12} + 
 \Delta_{22}\right] \nonumber\\
&\simeq& - \frac{N_c\Lambda^4 \Delta_{22} 
(\sigma_1 - \sqrt{3} \sigma_2)^2}{12\pi^2 m^2_{\sigma'} \sigma_1^2 
\ln\frac{\Lambda^2}{4\sigma^2_1}}  ;\nonumber\\
Z_{\sigma'} +  Z_{\sigma} &=& \frac{N_c\Lambda^4}{2\pi^2} \equiv Z_0\,. 
\label{scalcorr}
\end{eqnarray}
These relations are derived with the help of mass-gap eqs.(\ref{massgapQQM}).

To the first order in the leading-log approach, the weak decay
coupling constant for pion can be found from the gauged second variation:
\begin{equation}
F_{\pi}^2 \simeq \frac{N_c \sigma_1^2}{\pi^2}
\ln\frac{\Lambda^2}{4\sigma^2_1},
\end{equation}
and it coincides with that one of the NJL model.
Respectively the value of quark condensate  can be expressed in terms
of v.e.v. of  $\sigma_i$,
\begin{equation}
<\bar q q > \simeq - \frac{N_c\Lambda^2}{8\pi^2}
(\sigma_1 - \sqrt{3} \sigma_2).
\end{equation}
Thus taking these equations into account and remembering the leading order of
the $\pi'$-meson mass (\ref{masrelQQM}) one arrives to the remarkable relation:
\begin{equation}
Z_{\sigma} \simeq 4 \frac{<\bar q q >^2}{F_{\pi}^2}. \label{pcac}
\end{equation}
Let us examine the two-point correlator of local densities (\ref{localden}) 
in the pseudoscalar channel:
\begin{eqnarray}
\Pi_\pi (p^2) &=& - \frac{N_c\Lambda^2}{2\pi^2} + 
\frac{Z_\pi}{p^2} + \frac{Z_{\pi'}}{p^2 +m^2_{\pi'}} \,;\nonumber\\ 
Z_\pi &\simeq& Z_{\sigma} \quad \mbox{\rm for} \quad 
m^2_{\pi'} \simeq m^2_{\sigma'} ;\nonumber\\ 
Z_{\pi'} +  Z_{\pi} &= & Z_0\,;\qquad
Z_{\pi'}\simeq Z_{\sigma'} . \label{pseudocorr}
\end{eqnarray}
The equality of $Z_\pi$ and $Z_{\sigma}$ in eq.(\ref{pcac}) realizes
both the approximate restoration of chiral symmetry in each multiplet and
the fulfillment of PCAC requirement (\ref{pcac}) 
for the residue in the pion pole.
 
We stress that the residues in poles are of different order of magnitude: 
\begin{equation}
Z_{\sigma} \sim Z_\pi ={\cal O} 
\left({Z_0 \over \ln\frac{\Lambda^2}{4\sigma^2_1}}\right) 
\ll Z_{\sigma'} 
\sim Z_{\pi'} = {\cal O} \left(Z_0\right)\,. 
\end{equation}
Now we are able to impose and check the CSR constraints (\ref{constraints}).
The leading
asymptotics represents the generalized $\sigma$-model relation and is 
automatically fulfilled:
\begin{equation}
Z_{\sigma} + Z_{\sigma'} = Z_{\pi} + Z_{\pi'} = Z_0\,, \label{zeta}
\end{equation}
that in fact reflects  the manifest chiral symmetry of the QQM lagrangian.

As to the second constraint the possibility to satisfy it depends on the 
value of the QCD coupling constants $\alpha_s$. Indeed, it
can be written by means of (\ref{masrelQQM}) as follows:
\begin{eqnarray}
m_{\sigma'}^2 -  m_{\pi'}^2 &\simeq& \frac{\Delta_{SP}}{Z_0};\nonumber\\
\sigma^{2}_{1}
+\frac{2 \sqrt{3}}{3}\sigma_{1}\sigma_{2}
+ 3 \sigma^{2}_{2} &\simeq& \frac{3N_c\alpha_s }{8\pi} 
(\sigma_{1} - \sqrt{3} \sigma_{2})^2 .
\end{eqnarray} 
As the left part is always positive there exists a lower bound for 
$\alpha_s \geq \frac{8\pi}{9 N_c}$  providing solutions 
of the constraint. The lowest value of $\alpha_s \simeq 0.9 $
is given by $\sigma_{1} = - \sqrt{3} \sigma_{2}$ and for these v.e.v.'s
one obtains the following splitting between the $\sigma'$- and 
$\pi'$-meson masses:
\begin{equation}
m_{\sigma'}^2 -  m_{\pi'}^2 \simeq 
\frac83 \sigma^{2}_{1} = \frac16 m^2_{\sigma};
\end{equation}
{\it i.e.} for $m_{\sigma} = 500 \div 600 MeV$ 
these masses practically coincide,
$m_{\sigma'} \simeq  m_{\pi'} = 1300 MeV$ and such a $\sigma'$-meson
may be identified \cite{partdata} with $f_0 (1300)$. 
The above value
of $\alpha_s$ lies in the region of rather strong coupling where 
next-to-leading corrections to the anomalous dimension
of four-quark operator in (\ref{ChirSymRes}) 
are not negligible, 
$\sim \frac{\alpha_s}{\pi} \sim 0.3$ and 
should be systematically taken into account to obtain 
a reasonable precision.
 However the very fact that one has to match to
QCD asymptotics at a scale $\mu \sim 600 MeV \sim m_\sigma$ 
lower than the masses of heavy resonances is not troublesome as it
relates just coefficients of $ 1 / p^2$ expansion irrespectively of how 
high is the momentum. On the other hand the matching should be performed 
in the region where one can neglect even more heavier resonances, i.e. 
at a scale $\leq 1$ GeV. 
\section*{CONCLUSIONS}
\noindent
1. \quad We have shown that the 
quasilocal quark models truncating (perturbative)
low-energy QCD effective action can serve to describe the
physics of heavy meson resonances. The matching to nonperturbative QCD
based on the chiral symmetry restoration at high energies improves 
the predictability of such models. QQM extend the NJL model and inevitably
contain a rather light scalar meson which is however not excluded by
the particle phenomenology \cite{partdata}. \\

\noindent
2. \quad The fast convergence in QQM of mass spectra and other characteristics
of heavy parity doublers entail their decoupling from the 
low-energy pion physics. For instance, let us calculate the dim-4 
chiral coupling constant \cite{gasserleut1985,bijnrafzheng},
\begin{eqnarray}
L_8 &=& \frac{F^4_\pi}{64 < \bar q q >^2} \left(\frac{Z_\sigma}{m^2_\sigma}
+ \frac{Z_{\sigma'}}{m^2_{\sigma'}}
- \frac{Z_{\pi'}}{m^2_{\pi'}}\right)\nonumber\\
 &\simeq& \frac{F^2_\pi}{16 m^2_\sigma} \left(1 - 
\frac{6 \alpha_s \pi F^2_\pi m^2_\sigma }{m^4_{\pi'}}\right).
\end{eqnarray}
The second term represents the net effect of heavy resonances after the
CSR constraints (\ref{constraints}) have been imposed. 
It is easy to find that its relative
contribution is less than 2\%. Therefore this constant is essentially
determined in QQM by the lightest scalar meson. Its value,
$L_8 = (0.9 \pm 0.4)\times 10^{-3}$ from 
\cite{gasserleut1985,ecker1995,pich1995} nearly 
accepts $m_{\sigma} \simeq 600 MeV$. \\

\noindent
3. \quad Some disadvantage of 
QQM as well as of the original NJL model
is that they presuppose the large, critical values of  
four-quark coupling constants which
is difficult to justify with perturbative calculations in QCD. As well
the CSR matching has to be performed at a scale where the QCD 
coupling constant is rather large and the perturbation theory is unreliable.
The Extended Chiral Quark Models 
\cite{andresp1998,andresp1999} seem to be free of these shortcomings
and are able to adjust the light scalar meson mass to be of order
1 GeV.  However the final choice between them may be done by the
fit of a larger variety of meson characteristics which is in progress.\\

\noindent 
4. \quad Let us comment the approximations used to
derive the meson characteristics: namely, the large $N_c$
and leading-log approximations. The first one is equivalent
\cite{hooft1974,witten1979} to the
neglect of
meson loops. The second one fits well the quarks
confinement  as
quark-antiquark threshold contributions are suppressed in two-point
functions in the leading
log approximation. The accuracy of
this approximation is controlled also  by the magnitudes of higher dimensional
operators neglected in QQM, {\it i.e.} by contributions of heavy
mass resonances not included into QQM. All these approximations 
are mutually consistent. In particular, in the effective action without 
gluons the quark confinement should be realized with the help of
an infinite number of quasilocal vertices with higher-order derivatives.
Then the imaginary part of quark loops can be compensated and 
their momentum dependence can eventually reproduce the infinite sum of meson 
resonances in 
the large-$N_c$ limit. If the effective action is truncated with a finite
number of vertices and thereby deals
with only a few resonances  one has to retain only a finite number of
terms in the low-momentum expansion of quark loops in the CSB phase,
with a non-zero dynamical mass. \\

\noindent
5.\quad There are more possibilities to bind the phenomenological constants
of QQM based on CSR constraints for three- and four-point correlators and
also in the vector and axial-vector channels. Some of these CSR rules 
have been explored in 
\cite{andresp1998,donoghue1994,moussal1997,knechtrafael,perisperrot,knechtperis1998}.

\section*{ACKNOWLEDGEMENTS}

We express our gratitude to the organizers of the
International Workshop on Hadron Physics 1999 in Coimbra and
especially to Prof. J. da Providencia for
hospitality and financial support of one of us (V.A.).
We are also  grateful to D. Espriu and R. Tarrach for useful 
discussion and attention to our work.

\end{document}